# Tuning the Electronic Structure of Monolayer Graphene/MoS$_2$ van der Waals Heterostructures via Interlayer Twist

Wencan Jin,[1] Po-Chun Yeh,[1] Nader Zaki,[1] Daniel Chenet,[1] Ghidewon Arefe,[1]

Yufeng Hao,[1] Alessandro Sala,[2] Tevfik Onur Mentes,[2] Jerry I. Dadap,[1]

Andrea Locatelli,[2] James Hone,[1] and Richard M. Osgood, Jr.[1,*]

[1] *Columbia University, New York, New York 10027, USA*
[2] *Elettra-Sincrotrone Trieste S.C.p.A., I-34012, Basovizza, Trieste, Italy*

**We directly measure the electronic structure of twisted graphene/MoS$_2$ van der Waals heterostructures, in which both graphene and MoS$_2$ are monolayers. We use cathode lens microscopy and microprobe angle-resolved photoemission spectroscopy measurements to image the surface, determine twist angle, and map the electronic structure of these artificial heterostructures. For monolayer graphene on monolayer MoS$_2$, the resulting band structure reveals the absence of hybridization between the graphene and MoS$_2$ electronic states. Further, the graphene-derived electronic structure in the heterostructures remains essentially intact, irrespective of the twist angle between the two materials. In contrast, however, the electronic structure associated with the MoS$_2$ layer is found to be *twist-angle dependent*; in particular, the *relative* difference in the energy of the valence band maximum at $\bar{\Gamma}$ and $\bar{K}$ of the MoS$_2$ layer varies from approximately 0 to 0.2 eV. Our results suggest that monolayer MoS$_2$ *within the heterostructure* becomes predominantly an indirect bandgap system for all twist angles except in the proximity of 30 degrees. This result enables potential bandgap engineering in van der Waals heterostructures comprised of monolayer structures.**

The interest in two-dimensional (2D) materials and materials physics has grown dramatically over the past decade. The family of 2D materials, which includes graphene (Gr), transition metal dichalcogenides (TMDCs), hexagonal boron nitride (hBN), *etc*., can be fabricated into atomically thin films since the intralayer bonding arises from their strong covalent character, while the interlayer interaction is mediated by weak van der Waals forces. In addition to homogenous 2D materials, van der Waals (vdW) *heterostructures* [1] have recently emerged as a novel class of materials, in which different 2D atomic planes are vertically stacked to give rise to distinctive properties and exhibit new structural, chemical, and electronic phenomena [2-8]. These artificial heterostructures, in contrast with traditional heterostructures, can be designed and assembled by stacking individual 2D layers without lattice parameter constraints. The weak electron coupling at the interface of vdW heterostructures offers the possibility of combining the intrinsic electronic properties of the individual 2D layers. In particular, Gr/MoS$_2$ vdW heterostructures are remarkable because of the high carrier mobility [9] and broadband absorption [10] of graphene, as well as the direct bandgap [11-13] and extremely strong light-matter interactions [14] of monolayer MoS$_2$. The combination of these unusual characteristics has led to potential applications in field-effect transistor devices [15, 16], energy harvesting materials [17, 18], and memory cells [19, 20]. Despite this weak coupling, however, there is also the possibility of engendering emergent properties that are distinct from that of their constituent materials as has been observed in TMDC, e.g., the direct to indirect gap transition in going from monolayer to multilayer crystals. In fact, for the Gr/MoS$_2$ interface, density functional theory (DFT) calculations have predicted the crossover between a direct and indirect bandgap of MoS$_2$ induced by the modification of interlayer orientation [21, 22]. Thus, changing the relative orientation of the constituent gapless and direct-gapped 2D monolayers forming the heterostructure, results in the tunability of its the electronic structure. This tunability is of pervasive importance to the development of new high performance electronic devices. Also, the significant quenching photoluminescence peak intensity in Gr/TMDC heterostructures suggests a charge transfer between Gr and TMDC layer [18, 23]. These theoretical and optical investigations have led to a pressing need for a full understanding of the electronic structure of Gr/MoS$_2$ vdW heterostructures. Very recently, photoemission measurements of Gr/MoS$_2$ interface have been attempted [24-26]. Thus Coy-Diaz *et al.* examined a twisted interface between polycrystalline graphene and a *bulk* MoS$_2$ crystal [24, 25], while Miwa *et al.* examined the electronic structure of a multidomain epitaxial MoS$_2$-graphene heterostructure, which is laterally averaged over different orientations [26]. In all, a direct experimental investigation of the evolution of the momentum-resolved electronic structure with *twist angle* in Gr/MoS$_2$ twisted bilayer has, thus far, been lacking.

In this paper, we report direct measurement of the twist-angle-dependence of the local electronic structure of Gr/MoS$_2$ vdW heterostructures supported on a Si substrate with native oxide. In order to characterize our samples, we employed a range of methods based on synchrotron-based cathode lens microscopy, warranting high sensitivity to both the crystal and electronic structure. An important finding, using microprobe low energy electron diffraction (μ-LEED) line profile analysis, is that the corrugation of the graphene overlayer on MoS$_2$ is less than that of graphene on SiO$_2$. Selected-area angle-resolved photoemission spectroscopy (μ-ARPES) measurements show that the Dirac point is consistently located within experimental error at the Fermi level and that the Fermi velocity is close to that of pristine graphene, thus indicating that graphene remains essentially intact when placed on monolayer MoS$_2$ regardless of twist angle. The ARPES band maps of MoS$_2$ show the absence of band hybridization with Gr bands for occupied states within 2.5eV of the valence band maximum (VBM), but do show a dependence of the relative energy positions of the VBM at $\bar{K}$ and $\bar{\Gamma}$ on twist angle. Perhaps most significantly, our results suggest that monolayer MoS$_2$ within the heterostructure is predominantly an indirect bandgap system for all twist angles except at or near the twist angle of 30°.

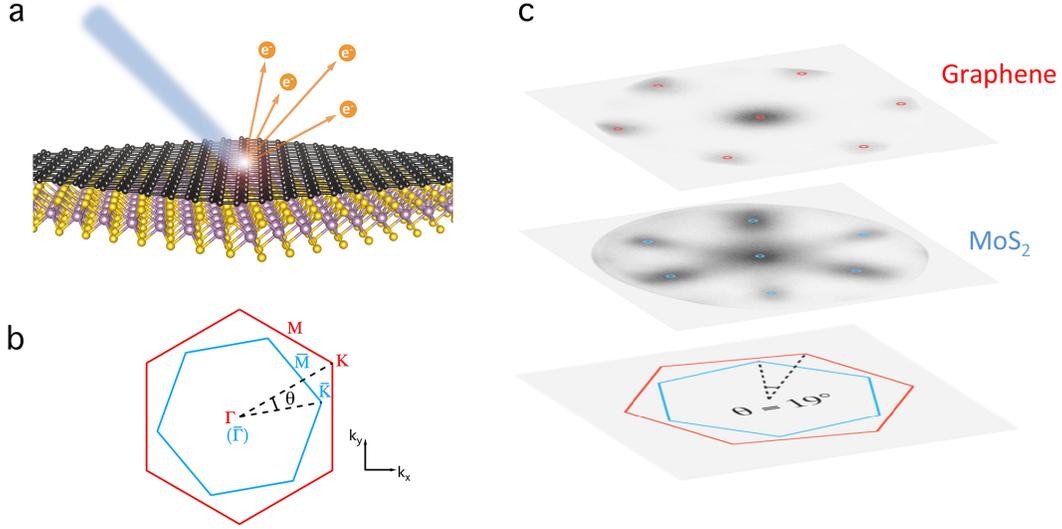

**Figure 1. a.** Schematic of the photoemission process and configuration. **b.** Brillouin zone (BZ) of Gr and surface Brillouin zone (SBZ) of MoS$_2$ with a twist angle of θ. We define the high-symmetry points of Gr BZ (red) as M-Γ-K and those of MoS$_2$ SBZ (blue) as $\bar{M}$-$\bar{\Gamma}$-$\bar{K}$. **c.** LEED patterns (upper plane) derived mostly from the graphene overlayer at 40 eV, and LEED pattern (middle plane) derived mostly from the MoS$_2$ bottom-layer at 45 eV. The diffraction spots are projected to the bottom plane to extract the twist angle.

Our measurements were performed on the spectroscopic photoemission and low energy electron microscopy (SPELEEM) in operation at the Nanospectroscopy beamline at the Elettra Synchrotron in Trieste, Italy [27, 28]. The μ-LEED measurements were restricted to regions of 1 and 0.5 μm in diameter. The μ-ARPES measurements were carried out with an energy resolution of 250 meV, at incident photon energy of 26 eV (see Supplementary Section 1 [29]).

Figure 1a shows the photoemission process and its configuration. The incident photon beam makes a 16° grazing angle with respect to the sample, leading to preferential probing of states derived from out-of-plane orbitals. We fabricated our samples by subsequent transfer of CVD-grown *monolayer* Gr [30] and CVD-grown *monolayer* MoS$_2$ [31] onto a *n*-doped Si(100) substrate with a native-oxide surface layer. As indicated in Ref. [31], this type of CVD-grown MoS$_2$ was carefully characterized using TEM, Raman, and photoluminescence, and was confirmed to be a uniform monolayer except for small multilayer patches in the center of the island. Due to the growth process, the Gr and $MoS_2$ domains are randomly rotated by a certain twist angle ($\theta$). Accordingly, the reciprocal space structures (Fig. 1b) are rotated by the same angle. This fact allows us to use $\mu$-LEED to determine the twist angle. In Fig. 1c, the stack shows the $\mu$-LEED images of Gr over $MoS_2$ with a finite twist angle. Using 40 eV incident electron beam, we obtain the diffraction pattern of the Gr overlayer and from which we see a six-fold-symmetry structure, as shown in the top plane of the stack. For the middle plane of the stack, on the other hand, an electron energy of 45 eV is used, for which the $\mu$-LEED pattern is from an exposed region of the bottom $MoS_2$ layer. Using 2D Gaussian fitting, we were able to determine the centers of the diffraction spots, which are denoted by colored circles in the LEED pattern. By projecting the two hexagonal spot arrays for Gr (red) and $MoS_2$ (blue) to the bottom plane of the stack, we obtain the twist angle $\theta$. Besides LEED measurements, twist-angle determination was carried out using ARPES constant energy maps (Supplementary Figure S3), which revealed agreement between these two methods.

LEED I-V measurements (tuning the incident electron beam energy from 20 to 100 eV) do not show evidence of any moiré structure, or spots arising from multiple scattering between the Gr and $MoS_2$ lattices, indicating a weak superlattice potential and lack of long-range coherence. While the ability to see a moiré structure can be hindered by spot broadening in LEED, we also do not see evidence of a superlattice potential in the ARPES measurements (discussed below), thus supporting the above claim.

Our previous work has shown that the width of the LEED (00) spot can be used as a signature of the corrugation of 2D materials [32, 33]. This approach to linewidth analysis has been used with the present LEED I-V measurements as well (see Supplementary Section 4). These measurements show that Gr on $MoS_2$ has a linewidth-derived angle variation of 4.1°±0.4°, which is less than the value of 6.1°±0.5° found in the case of graphene on $SiO_2$ [32]. This suggests that Gr on $MoS_2$ is less corrugated than on the widely used $SiO_2$ substrate.

After characterizing the crystal quality, we measured the electronic structure of the heterostructures using the ARPES capability of the SPELEEM system. Figure 2a shows the constant-energy map (CEM) of a graphene overlayer heterostructure at a binding energy of 875 meV. The contour is from the spectrum of the graphene-derived Dirac cones. Figures 2b show the ARPES bandmap and the corresponding second derivative intensity plot [34] of the graphene derived Dirac cone along the Γ-K direction. Note that there are no replica cones near the K or K′ points, which is also evidence for the absence of a moiré structure. Figures 2a-2b are acquired from the Gr overlayer with a twist angle of 19° with respect to the MoS$_2$ bottom-layer. The anisotropy of the spectral intensity in the CEM as well as in the ARPES bandmap is due to the photoemission selection rules [35]. From the data in Fig. 2b, we determine that the Dirac point resides in close vicinity of the Fermi level at the K point (the Dirac point is determined using MDCs fitting, and the Fermi level is determined using Fermi-function fitting; see supplementary Fig. S5). By fitting the band dispersion with a straight line, we obtain a Fermi velocity of $(0.99\pm0.01)\times10^6$ m/s, which is close to the value of pristine graphene [36]. We also investigated the band structure of Gr overlayer heterostructures for different twist angles. Figures 2c & 2d show the ARPES bandmap and the corresponding second-derivative-intensity plot of Gr with a twist angle of 12° and 28°, respectively, and the Fermi velocities that we extract for these two cases are $(0.96\pm0.02)\times10^6$ m/s and $(0.97\pm0.02)\times10^6$ m/s, respectively. Thus, within our energy and momentum resolution, we do not see significant electronic-structure changes of the graphene-derived bands with twist angle. Therefore, the electronic structure of monolayer graphene is essentially intrinsic when it is an overlayer on MoS$_2$, regardless of the twist angle. Based on the LEED intensity profile linewidth analysis and the ARPES band map, we conclude that monolayer MoS$_2$ is an ideal substrate for preserving the intrinsic properties of monolayer graphene.

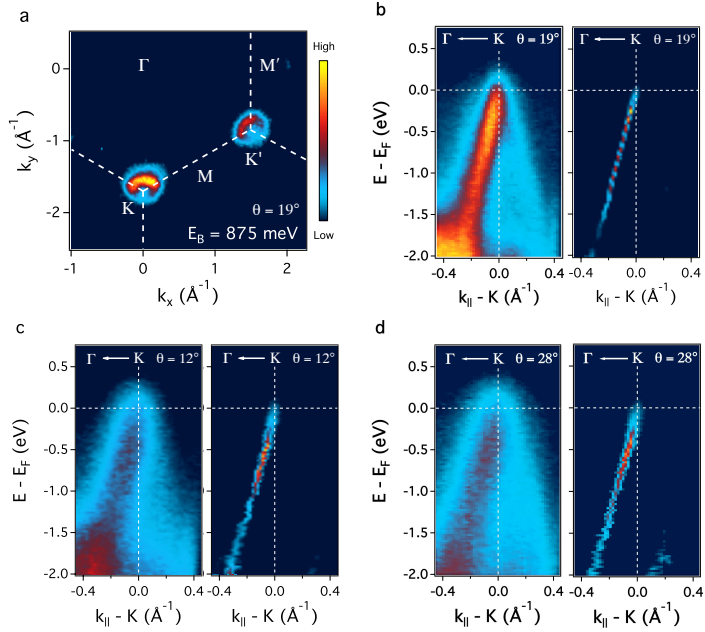

**Figure 2. a.** Constant energy map of a graphene overlayer heterostructure at a binding energy of 875 meV. **b.** ARPES band map (left) and second derivative intensity plot (right) of the graphene derived Dirac cone along Γ-K direction. **a-b** are acquired from the Gr overlayer with a twist angle of 19° with respect to MoS$_2$ bottom layer. **c & d.** ARPES bandmap and corresponding second derivative intensity plot of a Gr overlayer with a twist angle of 12° and 28°, respectively.

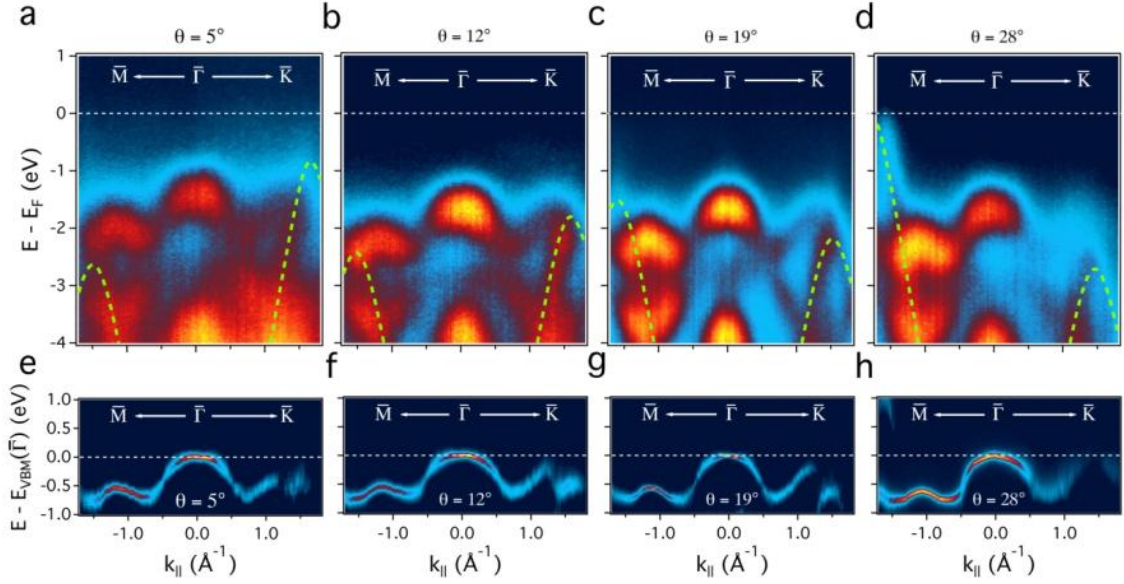

**Figure 3. a-d.** ARPES band map along $\bar{M}$-$\bar{\Gamma}$-$\bar{K}$ of the MoS$_2$ layer with a twist angle of 5°, 12°, 19°, and 28°, respectively. The green dashed curves are Gr derived band in a heterostructures acquired from a tight-binding model. **e-h.** Second derivative plot of the uppermost valence band in **a-d**.

It is known that the alignment of the energy bands at the interface significantly affects the behavior of semiconductor heterostructures [37]. ARPES allows us to obtain the band alignment between the Gr-derived bands and the MoS$_2$-derived bands directly. Thus, we find that for all our measured twist angles, Gr derived bands are very close to intrinsic and that the Gr-derived Dirac point is situated within the MoS$_2$ bandgap (Supplementary Figure S6).

In Ref. [24], the Ultraviolet Photoemission Spectroscopy (UPS) measurement of Gr capped bulk MoS$_2$ (with one particular relative rotation of 12°) shows a ~ 0.1 eV VBM shift in comparison with a bare bulk MoS$_2$ crystal, which hints at an electronic structure modification of MoS$_2$ in a Gr/MoS$_2$ interface. We thus measure the electronic band structure derived from the MoS$_2$ bottom-layer in the Gr/MoS$_2$ heterostructure. Figure 3a-3d shows the ARPES band maps along $\bar{M}$-$\bar{\Gamma}$-$\bar{K}$ of the MoS$_2$ SBZ for twist angles of 5°, 12°, 19°, and 28°, respectively. Besides MoS$_2$ derived bands, we also observe the overlay of Gr derived bands. To make a comparison, we use a nearest-neighbor tight-binding (NNTB) model [36] to generate the band dispersion of intrinsic monolayer Gr and superimpose these bands (green dashed curves) for specific twist angles onto the corresponding ARPES band maps. As shown in Figs. 3a-3d, the measured graphene-derived bands agree well with the NNTB bands of intrinsic graphene. Note that for all measured twist angles, there is no indication of band hybridization for Gr in the range of binding energies measured in this study, which is in good agreement with theoretical predictions [22]. In the corresponding momentum distribution curve (MDC) plot and second derivative plot, we can confirm the absence of hybridization (Supplementary Figure S7). Another set of ARPES measurements of an MBE-grown MoSe$_2$ thin film, which was formed on a bilayer of graphene/SiC, also showed no evidence of band hybridization between the MoSe$_2$ and graphene electronic states [38]. However, a recent ARPES study of CVD-grown graphene on a *bulk* MoS$_2$ crystal shows modification of the graphene π-bands by way of hybridization with bulk MoS$_2$ bands, mostly at higher binding energies than measured here [25]. Presumably, the increase in the number of states with out-of-plane character, as is the case for *bulk* MoS$_2$, increases the possibility for hybridization in comparison to our case of monolayer MoS$_2$.

Figures 3e-3h show the corresponding second-derivative intensity plots of the uppermost valence band (UVB) derived from MoS$_2$ as shown in Figs. 3a-3d. The intensity of the signal is strong in the $\overline{\Gamma M}$ direction but weak in the $\overline{\Gamma K}$ direction due to photoemission selection rules. The $\overline{\Gamma M}$ region is dominated by out-of plane Mo $d_{z^2}$ orbitals, while, in the vicinity of the $\overline{K}$ point, it is derived mainly from the in-plane Mo $d_{x^2-y^2}/d_{xy}$ orbitals [39]. In Fig. 3h ($\theta = 28°$), we find that the VBM at $\overline{K}$ and $\overline{\Gamma}$ are almost degenerate. However, for smaller twist angles and as shown in Figs. 3e-3g, the VBM at $\overline{K}$ is lower than that at $\overline{\Gamma}$. These results indicate that the relative position of the VBM of $\overline{\Gamma}$ and $\overline{K}$ is tuned by the twist angle. For comparison, we also have measured the heterostructure, in which MoS$_2$ is the overlayer (i.e., MoS$_2$/Gr where MoS$_2$ is on top) for the case of a 12° twist angle (Supplementary Figure S8). Note that the energy difference between the VBM of

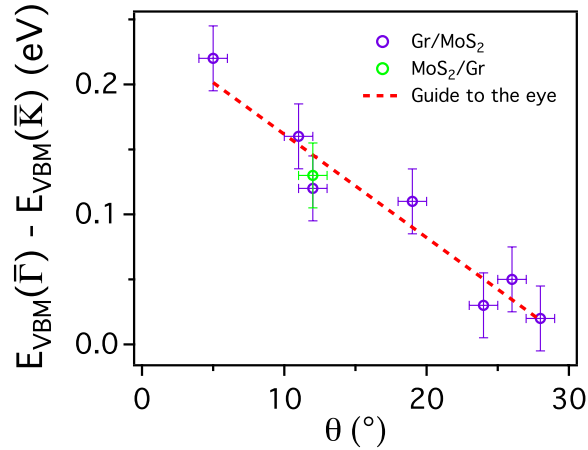

**Figure 4.** Energy difference between $\overline{\Gamma}$ and $\overline{K}$ *versus* twist angle in the Gr/MoS$_2$ (purple) and MoS$_2$/Gr (green) heterostructures.

$\bar{\Gamma}$ and $\bar{K}$ in this MoS$_2$/Gr heterostructure (0.13±0.03 eV) at this twist angle is almost identical to that of the Gr/MoS$_2$ heterostructure (0.12±0.03 eV) with the same twist angle, and both are close to the VBM shift (~0.1eV) in a Gr/Bulk MoS$_2$ interface [24].

In Fig. 4, we show the evolution of energy difference between the VBM of $\bar{\Gamma}$ and $\bar{K}$ with twist angle. The energy difference between the VBM of $\bar{\Gamma}$ and $\bar{K}$ is determined by using the energy distribution curve (EDC) peak fitting method (Supplementary Figure S9). The red dashed line is a guide to the eye to illustrate the overall trend in the data. We find that the energy difference appears to decrease gradually from ~0.2 eV to ~0 eV as the twist angle evolves from 5° to 28°. Thus, our results suggest that monolayer MoS$_2$ within the heterostructure is predominantly an indirect bandgap system for all twist angles except at or near the twist angle of 30°.

A detailed theoretical investigation is beyond the scope of this experimental paper; we thus discuss the physical origin of the electronic structure modification in Gr/MoS$_2$ vdW heterostructures in light of theoretical works already in the literature. In the first, Ebnonnasir *et al*. found that the tunable band structure of a Gr/MoS$_2$ heterostructure arises from twist-angle dependent strain, and specifically discussed two extreme cases, 0° and 30° [21]. For a 0° twist angle, charge loss affects the Mo-S bond length; for a 30° twist angle, on the other hand, it is predicted that charge loss of the Mo-S bond is absent because graphene has a different registry with respect to the S atoms for this orientation [21]. Note that comprehensive DFT calculations have shown that band structure of monolayer MoS$_2$ is significantly affected by bond length variation [40]. Similarly, Wang *et al*. attribute the twist-angle dependence of the MoS$_2$ band structure to strain, and they too mention the presence of charge redistribution at the interface [22]. Note, however, that these two predictions for the trend in the direct-to-indirect bandgap transition with twist angle differ qualitatively; the reason for this difference between these two theory reports is beyond the scope of this paper.

Our measurements show a trend in the direct-to-indirect bandgap that is similar to that predicted by Ebnonnasir *et al*., except that the angle assignment is reversed to that of their report; our $\bar{\Gamma}$-$\bar{K}$ trend is plotted in Fig. 4. While the reason for this inconsistency is not clear, note that these DFT calculations mentioned above assumed commensurability between Gr and MoS$_2$, which may not exactly be the case in experiment. Also, we note the presence of the *n*-doped Si substrate in our experiment which is not taken into account in either of the theoretical reports; the effect of the Si substrate on MoS$_2$, however, is expected to be weak based on a previous report [11, 13].

An important question is why a dramatic Fermi-level shift in Gr was not observed given that there may be charge transfer at the interface between MoS$_2$ and Gr. Quantitatively, the charge transfer amount in Ref. [22] is ~$10^{-4}$ electron per carbon (e/C). The position of the Fermi level of Gr can be estimated using $E_F \cong v_F\sqrt{|n|}$ [36], where $v_F$ is the Fermi velocity, and $n$ is the carrier concentration in Gr. A charge transfer of $10^{-4}$ e/C gives a ~60 meV Fermi level shift which is smaller than our energy resolution. Thus, in summary, the negligible Fermi level shift in Gr, observed here, is explainable by a modest charge transfer from MoS$_2$ to Gr.

In conclusion, our experiments have enabled us to directly measure the electronic structure of Gr/MoS$_2$ vdW heterostructures with different twist angles. We find that the Gr layer behaves as pristine graphene when transferred atop monolayer MoS$_2$ regardless of twist angle, and its Dirac point is situated within the bandgap of MoS$_2$. In contrast, the electronic structure associated with the MoS$_2$ shows obvious twist-angle-dependence, specifically a VBM shift between $\bar{\Gamma}$ and $\bar{K}$, a phenomenon that appears, as a result of calculations, to be a result of charge-transfer-induced strain. Our results further reveal a sufficiently weak superlattice potential between the Gr and MoS$_2$ layers, and that the Gr layer is relatively flat on top of the MoS$_2$, in contrast to a Gr/SiO$_2$ system. This work opens up one possible route to new designer heterostructures, which combine the relativistic Dirac Fermions in monolayer Gr with the twist-angle-tuned bandgap in monolayer TMDCs

## Supplementary materials

Atomic photoionization cross section; bright- and dark-field LEEM image of Gr/MoS$_2$; determination of twist angle with ARPES; determination of the mosaic spread of Gr/MoS$_2$; determination of the doping level of Gr using MDCs fitting; alignment of graphene π-bands and MoS$_2$ bands; absence of band hybridization confirmed by MDCs and second-derivative plot; electronic structure of MoS$_2$/Gr heterostructure; determination of the energy difference between VBMs using EDC peak fitting.

## Corresponding Author

* Email: osgood@columbia.edu

**Acknowledgements**

This work was supported by the Department of Energy, Office of Basic Energy Sciences, Division of Materials Sciences and Engineering under Award Contract No. DE-FG 02-04-ER-46157. D.C., G.A., Y.H., and J.H. were supported as part of the NSF MRSEC program through Columbia in the Center for Precision Assembly of Superstratic and Superatomic Solid (MDR-1420634).

# Supplementary Materials

**Table of contents**

1. **Atomic photoionization Cross Section**

As shown in Fig. S1, the atomic photoionization cross section for Mo 4$d$, S 3$p$ and C 2$p$ subshells as a function of photon energy [28] demonstrates that the incident photon energy of 26 eV is near the Cooper minimum of the S 3$p$ orbital. Therefore, the states probed in this measurement are derived primarily from the Mo 4$d$ and C 2$p$ orbitals.

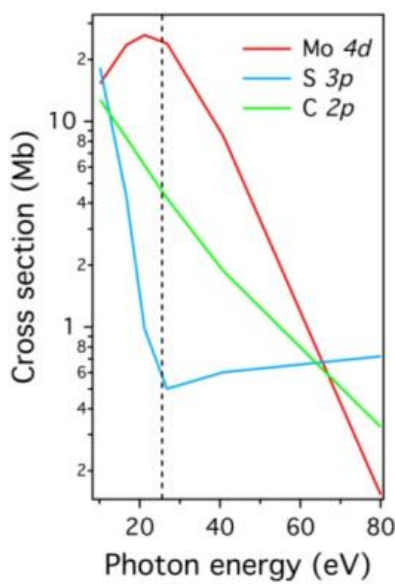

**Figure S1.** Atomic photoionization cross section for Mo 4$d$, S 3$p$ and C 2$p$ subshells as a function of photon energy. The black dashed line marks the photon energy of 26 eV.

## 2. Bright- and dark-field LEEM of Gr/MoS$_2$

Low energy electron microscopy (LEEM) is an atomic structural probe that enables direct imaging, with nanometer lateral resolution, the lateral extent of single-crystal samples. The selection of zero-order LEED beam produces image contrast known as bright field (BF) mode. The diffraction contrast can also be exploited in the dark field (DF) mode by imaging with a higher-order LEED beam. The BF LEEM image (Fig. S2a) shows that the graphene overlayer is uniform and flat. Figure S2b is the corresponding DF image, and its contrast allows visualization of the different domains of the graphene (Gr) overlayer. Thus, we have chosen *single-domain* regions to acquire µ-LEED and µ-ARPES data.

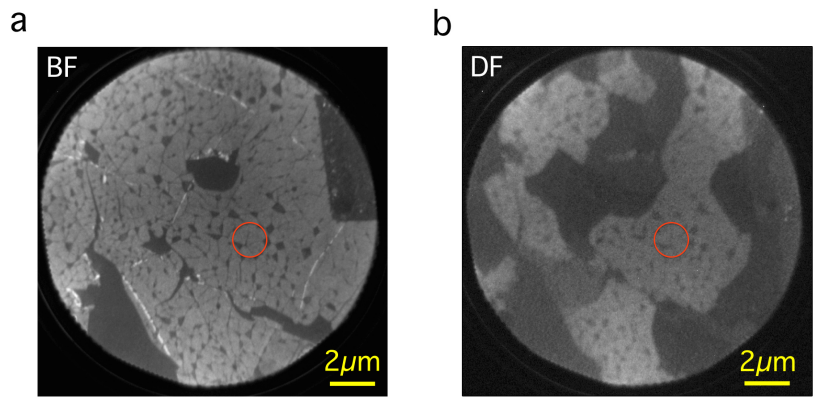

**Figure S2. a.** Bright field and **b.** Dark field LEEM image of a Gr/MoS$_2$ heterostructure. Red circles mark the spot where we acquire µ-LEED and µ-ARPES data.

## 3. Determination of twist angle with ARPES (Angle Resolved Photoemission Spectroscopy)

In our experiments, it is necessary to measure accurately the lattice twist angle between the Gr and MoS$_2$ monolayers. This measurement can be done with LEED or ARPES. In general, we found it more convenient to use ARPES data for both angle *and* band measurements. Thus we used the high symmetry features in ARPES constant energy maps (CEM). In Fig. S3, the upper plane of the stack shows the CEM near the Dirac point of Gr. The spectrum features show six-fold symmetry, and the red hexagon encloses the Brillouin zone (BZ) of Gr. The middle plane of the stack shows the CEM at the binding energy of the valence band maximum of $\overline{M}$ of MoS$_2$. The green hexagon denotes the BZ of MoS$_2$ and the white ovals enclose the features derived from $\overline{M}$ of MoS$_2$. The outer perimeter spectrum intensity in the middle plane comes from Gr. By superimposing the red and green hexagons together as shown in the bottom plane, we obtain the relative rotation angle between them, i.e., the twist angle of Gr and MoS$_2$. Note that the ARPES CEMs are acquired from the same sample that yields the LEED patterns shown in Fig. 1. Finally, the twist angle determined by ARPES CEMs is consistent with that obtained from the LEED measurements within an accuracy of 1°.

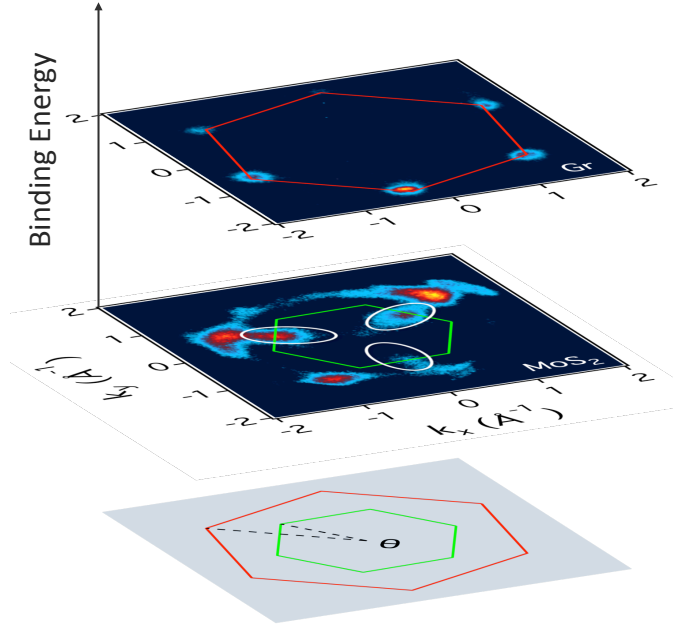

**Figure S3. (Upper plane)** Constant energy map of graphene in the vicinity of the Dirac points. The red hexagon denotes the Brillouin zone of graphene. **(Middle plane)** Constant energy map of MoS$_2$ at the binding energy of valence band maximum of $\bar{M}$. The green hexagon denotes the MoS$_2$ Brillouin zone and the white ovals encircle the spectrum features derived from $\bar{M}$ of MoS$_2$. **(Bottom plane)** Relative rotation angle between the red and green hexagon.

4. **Determination of the mosaic spread of Gr/MoS$_2$**

As shown in Fig. S4, the Gaussian linewidth of the (00) spots increases linearly with $k$, where $k$ is the total momentum of the LEED electrons, and is related to the incident electron energy according to $k = \sqrt{2m_e E_k}/\hbar$. The upper inset shows the diffraction pattern of the Gr overlayer for an incident electron energy of 45eV. The purple line across the (00) spot denotes the position where the line profile is acquired. The lower inset shows the line profile (purple curve) of the (00) spot and the corresponding Gaussian fitting (red curve). Assuming a Gaussian distribution for the local surface normal, the standard deviation, $\Delta\theta_{norm}$, can be obtained using a simple trigonometric relation

$$\Delta\theta_{norm} = \frac{1}{2}\frac{\hbar\Delta k_{\|}}{\sqrt{2m_e E_k}}$$

where $\Delta k_{\parallel}$ is the Gaussian width of the central diffraction maximum. Application of the above formula to the line in Fig. 3b results in a $\Delta\theta_{norm}$ of 4.1°±0.4°, which is smaller than the value of 6.1°±0.5° for the case when Gr is placed on a SiO$_2$ substrate [31].

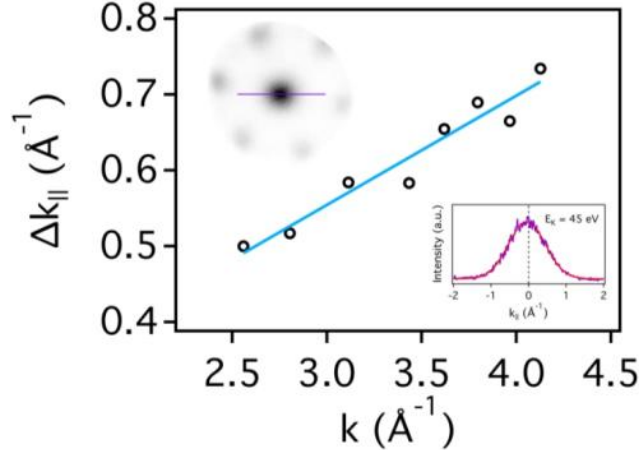

**Figure S4.** Gaussian Linewidth of the graphene (00) spot as a function of k. Black empty circles are the raw data and the blue curve is the linear fitting. Upper inset shows the LEED pattern acquired at 45 eV incident electrons. The purple line denotes the line profile across the (00) spot. Lower inset shows the corresponding line profile (purple) and Gaussian fit (red).

5. **Determination of the doping of Gr using MDCs fitting**

To determine the doping level, we perform (momentum distribution curve) MDC fitting. Figure S5a shows the ARPES band map of a graphene Dirac cone along the cut indicated in the inset. Figure S5b shows the corresponding MDCs plot. Figure S5c shows an example of the MDC fitting, in which the blue dots are from the raw data, and the red curve is the double Lorentzian fit to the raw data. These fittings are applied to the MDCs over a binding energy range of ~1-1.5eV and the MDC peaks are denoted as black dots in Figure S5a. Subsequently, we fit the MDC peaks using a straight line (dashed lines in Fig. 5S a-b). The place where these two straight lines intersect is denoted as the Dirac point ($E_D$). Figure S5d shows the integrated spectrum of the ARPES bandmap in Fig. S5a. We fit this cut-off feature to a Fermi-Dirac function with a linear background and find that the Fermi level ($E_F$, marked as green dashed line in Fig. S5d) appears to be 12±9 meV above the Dirac point. Since the energy difference between $E_D$ and $E_F$ is much smaller than our energy resolution, we presume that graphene is intrinsic or very close to intrinsic.

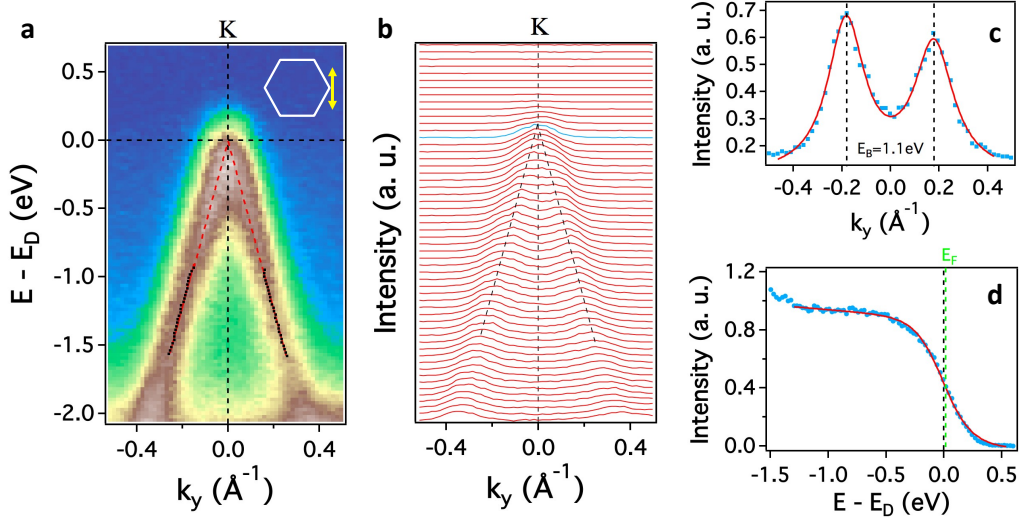

**Figure S5. a.** ARPES band map of graphene Dirac cone along the cut shown in inset. **b.** Corresponding MDCs plot. **c.** Double Lorentzian fit to the MDC at binding energy 1.1 eV. **d.** Integrated spectrum and corresponding Fermi-Dirac function fitting of ARPES bandmap shown in **a**.

### 6. Alignment of graphene π-bands and MoS$_2$ bands

As shown in Fig. S6 a-e, the Dirac cones of Gr at *K* and *K'* intersect the Fermi level at the Dirac point regardless of twist angle. The intense spectral feature, located over the range of binding energy from 1-2 eV, is the valence band maximum of MoS$_2$ at its BZ center ($\bar{\Gamma}$). Note that for all measured twist angles, no spectrum from the conduction band of Gr nor MoS$_2$ is present. Therefore, as indicated in Fig. S6f, Gr is intrinsically doped and its Dirac point is situated within the band gap of MoS$_2$.

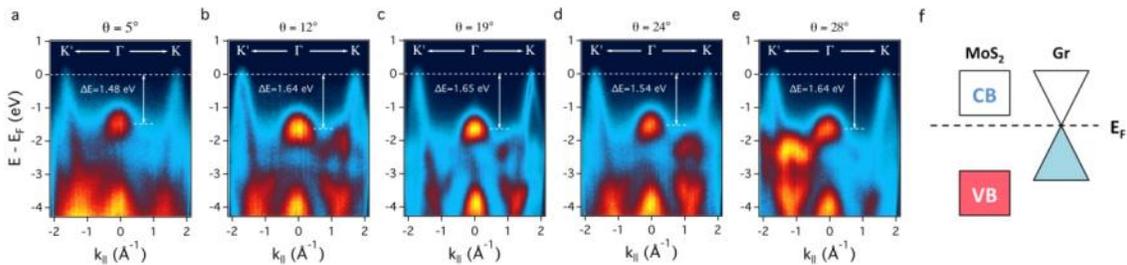

**Figure S6. a-e.** ARPES band maps along *K'-Γ-K* direction of the BZ of Gr with the twist angle of 5°, 12°, 19°, 24°, and 28°, respectively. **f.** Schematic of the band alignment between Gr and MoS$_2$.

## 7. Absence of band hybridization confirmed by MDCs and second-derivative plot

Figure S7a shows the MDC plot of the MoS$_2$ band with twist angle of 28° along the $\bar{M}$-$\bar{\Gamma}$ direction. Figure S7b presents the second derivative plot of the ARPES bandmap shown in Fig. 3d. As shown in Fig. S7, in the vicinity of $\bar{M}$, the Gr Dirac cone intersects the MoS$_2$ bands clearly without hybridization.

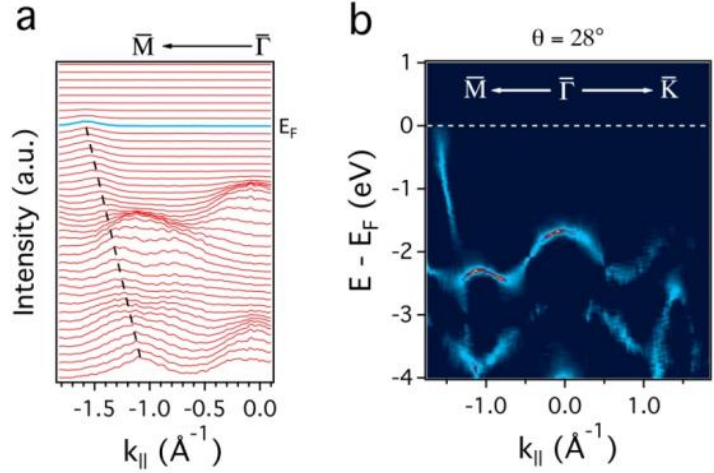

**Figure S7. a.** MDC plot of the ARPES band map along $\bar{\Gamma}$-$\bar{M}$ direction. The black dashed line is a guide to the eye of the graphene Dirac cone. **b.** Second derivative plot of ARPES band map.

## 8. Electronic structure of MoS$_2$/Gr heterostructure

Figure S8a shows an ARPES bandmap of the MoS$_2$ overlayer along $\bar{M}$-$\bar{\Gamma}$-$\bar{K}$. As expected, the spectral intensity of the heterostructure, where MoS$_2$ is the overlayer, is stronger than that for the case of the Gr/MoS$_2$ heterostructure, and the signal from the Gr bottom-layer, in this case, is relatively weak. Figure S8b shows the energy distribution curve (EDC) at $\bar{\Gamma}$ and $\bar{K}$. The red and blue dashed lines mark the peaks of the EDCs, i.e., the binding energies of the VBM at $\bar{\Gamma}$ and $\bar{K}$ shown in Fig. S8a. The VBM of $\bar{K}$ is 0.13±0.03 eV lower than that at $\bar{\Gamma}$. This value is confirmed in the second derivative intensity plot shown in Fig. S8c. Note that in Ref. [37] of the main text, Zhang *et al*. reported on the electronic structure of epitaxial MoSe$_2$ grown on bilayer graphene on SiC. Their LEED measurements show that epitaxially-grown MoSe$_2$ thin films are consistently aligned in the same lattice orientation (i.e. 0° twist angle) with the underlying bilayer graphene substrate. In that interface, in contrast to our measurement, both the MoSe$_2$ and bilayer graphene electronic structure are intact. This apparent discrepancy is probably due, first of all, to the different sample preparations and system configurations. Moreover, the structural relaxation used in the calculation in Ref. [37] allows an estimate of the interlayer distance between MoSe$_2$ and bilayer graphene to be ~4.2Å. However, in the two theoretical investigations discussed in the main text, the interlayer distance between MoS$_2$ and *monolayer* graphene is estimated to be 3.1Å [20] and 3.3Å [21], respectively. This result implies that the interlayer interaction of MoSe$_2$/bilayer Gr interface is apparently much weaker than that it is for the MoS$_2$/Gr interface.

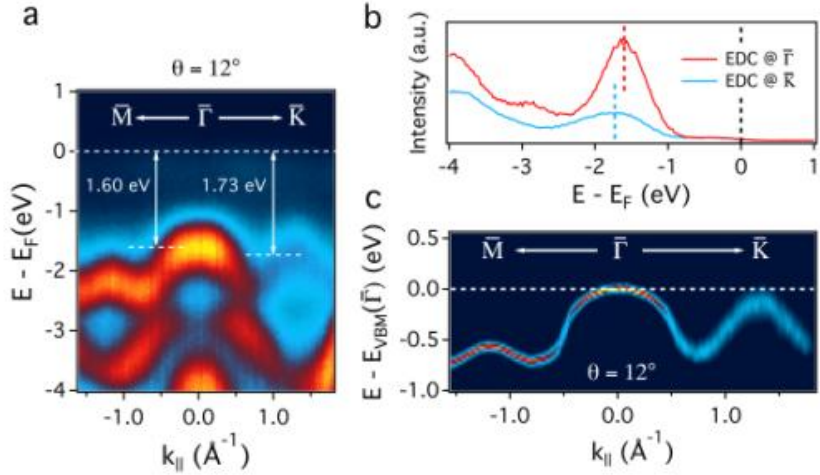

**Figure S8. a.** ARPES bandmap of MoS$_2$ along $\bar{M}$-$\bar{\Gamma}$-$\bar{K}$ in a MoS$_2$/Gr heterostructure with a twist angle of 12°; **b.** EDC at $\bar{\Gamma}$ (red) and $\bar{K}$ (blue); **c.** Second-derivative intensity plot of uppermost valence band shown in **a.**

9. **Determination of the energy difference between VBMs using EDC peak fitting**

To determine the energy difference between the valence band maximum (VBM) at $\bar{K}$ and $\bar{\Gamma}$, a curve fitting method to extract the peak positions of EDCs is applied. Figure S9 shows the peak fitting to the EDCs at $\bar{K}$ (upper panel) and $\bar{\Gamma}$ (lower panel) with twist angles of 5°, 12°, 19°, and 28°, respectively. To better visualize the peaks, we use a 4$^{th}$-order polynomial fit to extract the background (blue curves), and subtract the background from the raw EDC data (black dots). In Fig. S9a, a single Gaussian peak fitting is applied to the background-subtracted EDC. In Fig. S9 b-d, an extra peak derived from the Gr derived band emerges between a binding energy of 2-3 eV. Therefore, a double Gaussian peak fitting is utilized. Black and blue dashed lines mark the peak positions corresponding to the VBM at $\bar{K}$ and $\bar{\Gamma}$. Specifically, the VBM at $\bar{K}$ is 1.70 eV, 1.77 eV, 1.76 eV, and 1.67 eV for the twist angle of 5°, 12°, 19°, and 28°, respectively; the VBM at $\bar{\Gamma}$ is 1.48 eV, 1.64 eV, 1.65 eV, and 1.65 eV for the twist angle of 5°, 12°, 19°, and 28°, respectively.

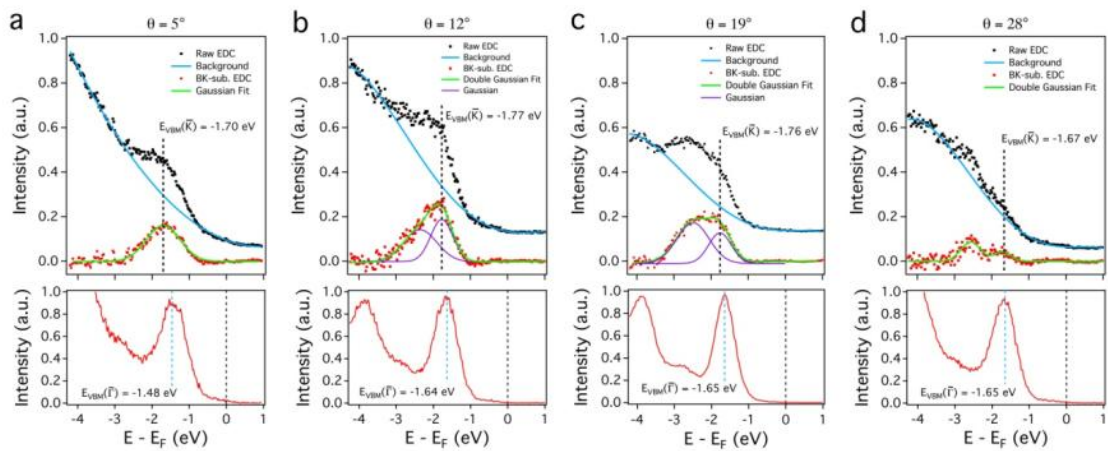